\documentclass[aps,prb,twocolumn,floatfix,superscriptaddress]{revtex4}
\usepackage[utf8]{inputenc}
\usepackage{mathrsfs}
\usepackage{graphicx}
\usepackage[english]{babel}
\usepackage{amsmath}
\usepackage{amssymb}
\usepackage{xcolor}
\usepackage{relsize}
\usepackage{CJK}
\usepackage{textcomp}
\usepackage{dsfont}
\usepackage{bm}

\newcommand{\me}{\mathrm{e}}
\newcommand{\mi}{\mathrm{i}}

\newcommand{\dif}{\mathrm{d}}

\allowdisplaybreaks
\begin{document}
	
\title{Thermal Uhlmann-Chern Number: Bridging Pure and Mixed States}

\author{Xin Wang}
\affiliation{School of Physics, Southeast University, Jiulonghu Campus, Nanjing 211189, China}

\author{Xu-Yang Hou}
\affiliation{School of Physics, Southeast University, Jiulonghu Campus, Nanjing 211189, China}



\author{Yan He}
\email{heyan_ctp@scu.edu.cn}
\affiliation{College of Physical Science and Technology, Sichuan University, Chengdu, Sichuan 610064, China}

\author{Hao Guo}
\email{guohao.ph@seu.edu.cn}
\affiliation{School of Physics, Southeast University, Jiulonghu Campus, Nanjing 211189, China}
\affiliation{Hefei National Laboratory, Hefei 230088, China}

\begin{abstract}
Topological properties of quantum systems at finite temperatures, described by mixed states, pose significant challenges due to the triviality of the Uhlmann bundle. We introduce the thermal Uhlmann-Chern number, a generalization of the Chern number, to characterize the topological properties of mixed states. By inserting the density matrix into the Chern character, we introduce the thermal Uhlmann-Chern number, a generalization of the Chern number that reduces to the pure-state value in the zero-temperature limit and vanishes at infinite temperature, providing a framework to study the temperature-dependent evolution of topological features in mixed states. We provide, for the first time, a rigorous mathematical proof that the first- and higher-order Uhlmann-Chern numbers converge to the corresponding Chern numbers in the zero-temperature limit, differing only by a factor of $1/D$ for $D$-fold degenerate ground states. We demonstrate the utility of this framework through applications to a two-level system, the coherent state model, the 2D Haldane model, and a four-band model, highlighting the temperature-dependent behavior of topological invariants. Our results establish a robust bridge between the topological properties of pure and mixed states, offering new insights into finite-temperature topological phases.
\end{abstract}

\maketitle

\section{Introduction}
The discovery of topological phases of matter, such as topological insulators and quantum Hall systems, has profoundly reshaped the landscape of condensed matter physics \cite{Hasan2010, Qi2011, shen2012topological, stanescu2024introduction, TKNN, Haldane, KaneRMP, ZhangSCRMP, KaneMele, KaneMele2, ChiuRMP, Bernevigbook, BernevigPRL, MoorePRB, FuLPRL, zak1989berry, atala2013direct}. Central to these phenomena is the concept of geometric phase, introduced by Berry in his seminal work \cite{Berry1984}. The Berry phase captures the phase accumulated by a quantum state during adiabatic and cyclic evolution in parameter space. It depends only on the path taken and reveals the geometric structure underlying the system's evolution.
Topological phases are characterized by robust invariants, such as the Chern number, which are derived from the Berry connection and curvature. These quantities describe the global topological properties of the ground-state wavefunction and are naturally formulated within the mathematical framework of principal bundles \cite{Nakahara2003, Budich2015, Guo20}. The Berry phase thus serves as a fundamental tool for understanding these invariants and the topological nature of quantum systems.

While the Berry phase successfully captures the topological properties of pure quantum states at zero temperature, most physical systems in practice exist at finite temperatures, where states are mixed and described by density matrices. To extend geometric phase concepts to such settings, Uhlmann introduced a generalized framework based on the parallel transport of density matrix purifications \cite{Uhlmann1986, Uhlmann1991}. The resulting Uhlmann phase provides a natural extension of the Berry phase to mixed states. Recent studies have shown that it can effectively capture finite-temperature topological phase transitions in a variety of systems \cite{Viyuela2014, Viyuela2014b, PhysRevLett.113.076407, OurPRA21, Morachis_Galindo_2021, OurPRB20b, Zhang21, PhysRevB.110.035144, PhysRevB.97.235141, OurUhlmannQuench}. A fundamental distinction between zero- and finite-temperature quantum systems lies in the topology of the associated fiber bundles: while the principal bundle for pure states is often topologically nontrivial, that for mixed states (called Uhlmann bundle for convenience) is always trivial, as the unique square root of the density matrix defines a global cross section \cite{Budich2015, Guo20}.
Nevertheless, several examples show that the Uhlmann phase reduces to the Berry phase in the zero-temperature limit for non-degenerate systems, suggesting that a deep correspondence may exist between the topologies of pure and mixed states \cite{Viyuela2014, Morachis_Galindo_2021, OurUB}. This potential relationship has been further substantiated through analytical and numerical studies under specific conditions \cite{OurDPUP, OurUB}. More recently, this connection has been extended to degenerate systems \cite{UWZ25}, revealing that, depending on the underlying topology, the Uhlmann phase may or may not coincide with the scalar Wilczek-Zee phase in the zero-temperature limit.
However, the topological triviality of the Uhlmann bundle precludes the direct definition of nontrivial Chern numbers, making it difficult to characterize topological properties at finite temperatures \cite{Viyuela2014, Budich2015}. Several proposals have attempted to address this issue, including the topological Uhlmann number \cite{Viyuela2014b, leonforte2019uhlmann} and approaches based on the gauge structure of spectrally constrained density matrices \cite{Budich2015}. However, these methods are generally less effective than the Chern number derived from the Berry curvature of pure states. Moreover, even the definition of the topological Uhlmann number suffers from a lack of uniqueness \cite{Budich2015}. Alternative frameworks have also been proposed for defining geometric phases in mixed states, such as the interferometric geometric phase \cite{PhysRevLett.85.2845}, which has attracted considerable attention \cite{PhysRevA.67.020101, PhysRevA.70.052109, Faria_2003, Chaturvedi2004, PhysRevLett.93.080405, Kwek_2006} due to its experimental accessibility \cite{PhysRevLett.91.100403, Ghosh_2006, PhysRevLett.101.150404, PhysRevLett.94.050401}. However, its topological interpretation remains incomplete, as a clear fiber-bundle formulation is still lacking. For this reason, it lies outside the scope of the present work.

In this work, we do not attempt to directly construct topological invariants of the mixed-state principal bundle (the Uhlmann bundle). Instead, we aim to identify a quantity that can be directly connected to the Chern number of pure states. Inspired by the approach in Ref. \cite{PhysRevB.97.235141}, we introduce the thermal Uhlmann-Chern number, which is generally not a topological invariant and hence not quantized in most cases.
Importantly, it can be rigorously shown that in the zero-temperature limit, this quantity reduces to a Chern number of a certain order, with a scaling factor of
$1/D$ for systems possessing a $D$-fold degenerate ground state. Here, the ``certain order" refers to the fact that, especially for higher-dimensional quantum systems, topological properties may be characterized by higher-order Chern numbers, and our thermal Uhlmann-Chern number naturally recovers the corresponding order Chern number in the zero-temperature limit. This result establishes a precise correspondence between the topological features of pure and mixed states, bridging the conceptual gap between zero- and finite-temperature regimes.
We illustrate the utility of the thermal Uhlmann-Chern number through applications to various models, including a two-level system, a coherent-state model, the two-dimensional Haldane model, and a four-band model. In all cases, it successfully captures temperature-dependent topological behavior. Our findings provide a robust and physically meaningful framework for studying topological phases at finite temperatures, with potential implications for experimental investigations of mixed-state topological properties \cite{Bardyn2018}.

The rest of the paper is organized as follows. Section~\ref{Formalism} introduces some fundamental concepts such as Berry, Wilczek-Zee connections, Uhlmann connection and Uhlmann-Chern number. Section~\ref{fUCnum} provides concrete examples, including a general two level model, the 2D Haldane model, and a coherent state system, to testify the correspondence between Chern number and Uhlmann-Chern number in the zero-temperature limit. Then we present a mathematical proof of this correspondence in Section~\ref{prooffiniteD}. Section~\ref{secUCnum} examines a four level model with two doubly degenerate subspaces and gives the calculations of second-order thermal Uhlmann-Chern number, which reduces to half the second-order Chern number in the zero-temperature limit. Section~\ref{UWCcurv} proposes a general proof of this correspondence for systems with $D$-fold degenerate ground state, which can be viewed as an extension of Section~\ref{prooffiniteD}. Finally, Section~\ref{Sec.6} summarizes our findings and presents concluding remarks. The Appendix shows the traceless nature of the Uhlmann curvature.

\section{Basic Formalism}\label{Formalism}
\subsection{Berry, Wilczek-Zee Connections and Chern number}
The formalism of the Berry connection and its associated curvature originates from the theory of the Berry phase, which is the geometric phase acquired by a quantum state during an adiabatic, cyclic evolution in a parameter space $M$, parameterized by $\mathbf{R} = (R_1, R_2, \dots, R_k)^T$. For a state $|\psi(\mathbf{R})\rangle$, the Berry connection is defined as:
\begin{equation}
	\mathcal{A}_{\mathrm{B}} = \langle \psi(\mathbf{R}) | \mathrm{d} | \psi(\mathbf{R}) \rangle,
\end{equation}
where $\mathrm{d}$ is the exterior derivative on $M$, making $\mathcal{A}_{\mathrm{B}}$ a 1-form. This Abelian connection gives rise to the Berry curvature
\begin{equation}
	\mathcal{F}_{\mathrm{B}} = \mathrm{d} \mathcal{A}_{\mathrm{B}} = \mathrm{d} \langle \psi(\mathbf{R}) | \wedge \mathrm{d} | \psi(\mathbf{R}) \rangle.
\end{equation}
When $\dim M=2$, the (first-order) Chern number is given by
\begin{equation}\label{n1}
	n^{(1)}=\frac{\mi}{2\pi}\int_M 	\mathcal{F}_{\mathrm{B}}.
\end{equation}
This quantity characterizes the topological properties of a principal bundle whose base manifold is $\mathcal{H} = \{|\psi(\mathbf{R})\}$ and whose fiber is the structure group $U(1)$.

This framework extends naturally to degenerate systems. Consider a set of $D$ degenerate states $\{|\psi_a\rangle\}$, where $a = 1, 2, \dots, D$. The Wilczek-Zee connection is defined as:
\begin{align}
	\mathcal{A}^{\mathrm{WZ}}_{ab} = \langle \psi_a(\mathbf{R}) | \mathrm{d} | \psi_b(\mathbf{R}) \rangle.
\end{align}
This constitutes a non-abelian gauge connection. The corresponding Wilczek-Zee curvature is given by:
\begin{align}
	\mathcal{F}^{\mathrm{WZ}}_{ab} = \mathrm{d} \mathcal{A}^{\mathrm{WZ}}_{ab} + \mathcal{A}^{\mathrm{WZ}}_{ac} \wedge \mathcal{A}^{\mathrm{WZ}}_{cb},
\end{align}
where summation over repeated indices (e.g., $c$) follows the Einstein convention. For a parameter space with $\dim M = 2k$, the Chern number is generalized to its $k$-th order form
\begin{align}\label{nk}
	n^{(k)} = \frac{1}{k!} \left( \frac{\mathrm{i}}{2\pi} \right)^k \int_M \operatorname{Tr} ( \mathcal{F}^{\mathrm{WZ}} \wedge \cdots \wedge \mathcal{F}^{\mathrm{WZ}} ).
\end{align}

\subsection{Uhlmann Connection and thermal Uhlmann-Chern number}

To generalize the geometric phase to mixed states, consider a quantum system with an $N$-dimensional Hilbert space described by the density matrix $\rho$. A full-rank density matrix can be purified through an amplitude $W = \sqrt{\rho} \mathcal{U}$, where $\mathcal{U} \in U(N)$ is the unitary phase factor. The density matrix is recovered as $\rho = W W^\dagger$, independent of $\mathcal{U}$. Diagonalizing $\rho = \sum_n \lambda_n |n\rangle \langle n|$, the purification reads
$W = \sum_n \sqrt{\lambda_n} |n\rangle \langle n| \mathcal{U}$.
For a cyclic evolution $\rho(t) = \rho(\mathbf{R}(t))$ along a closed path $C\in M$, if $W(t)$ satisfies the Uhlmann parallel-transport condition, i.e.
$W^\dagger \dot{W} = \dot{W}^\dagger W$,
 $W(t)$ is said to be a horizontal lift of $\rho(t)$. This implies the differential equation
 \begin{align}
\dot{\mathcal{U}} = -\mathcal{A}_\mathrm{U} \mathcal{U},
 \end{align}
where $\mathcal{A}_\mathrm{U}$ is the Uhlmann connection.
In general, $W(t)$ is not closed, thereby the initial and final $\mathcal{U}$ differ by the Uhlmann holonomy
 \begin{align}
\mathcal{U}(\tau) = \mathcal{P} \exp\left(-\oint_C \mathcal{A}_\mathrm{U}\right) \mathcal{U}(0),
 \end{align}
with $\mathcal{P}$ being the path-ordering operator. The Uhlmann connection $\mathcal{A}_\mathrm{U}$ can be explicitly written as
 \begin{align}\label{ucon}
\mathcal{A}_\mathrm{U} = -\sum_{nm} |n\rangle \frac{\langle n| [\mathrm{d}\sqrt{\rho}, \sqrt{\rho}] | m \rangle}{\lambda_n + \lambda_m} \langle m|,
 \end{align}
 and the associated Uhlmann curvature is $\mathcal{F}_\text{U}=\dif \mathcal{A}_\mathrm{U}+\mathcal{A}_\mathrm{U}\wedge \mathcal{A}_\mathrm{U}$.
For finite-dimensional systems, it can be shown that $\mathcal{F}_\mathrm{U}$ is traceless (see Appendix~\ref{app1b}), leading to the vanishing of the first-order Chern number in two-dimensional parameter space
\begin{align}\label{ChU}
	\text{Ch}_{\mathrm{U}} = \frac{\mi}{2\pi}\int_M \text{Tr}\mathcal{F}_{\mathrm{U}} = 0,
\end{align}
consistent with the topological triviality of the Uhlmann bundle.

To recover nontrivial topological features at finite temperature, one may consider the thermal-weighted trace $\operatorname{Tr}(\rho \mathcal{F}_\mathrm{U})$, referred to as the thermal Uhlmann curvature, which incorporates contributions from $\sum_n \lambda_n \langle n| \mathcal{F}_\mathrm{U} | n \rangle$. For a thermal state $\rho = \frac{1}{Z} \me^{-\beta H}$, where $\beta = 1/T$ and $Z = \operatorname{Tr} \me^{-\beta H}$, only the ground state contributes in the zero-temperature limit, as $\lambda_0 \to 1$ and $\lambda_{n>0} \to 0$.
This motivates the definition of the first-order thermal Uhlmann-Chern number \cite{PhysRevB.97.235141}:
\begin{equation}\label{eq:Thermal_Ch}
n^{(1)}_{\mathrm{U}} = \frac{\mathrm{i}}{2 \pi} \int_M \operatorname{Tr} \left( \rho \mathcal{F}_{\mathrm{U}} \right),
\end{equation}
which may reduce to the zero-temperature Chern number in Eq.\eqref{n1}. For higher-dimensional parameter space $M$, this generalizes to the $k$-th thermal Uhlmann-Chern number:
\begin{equation}
n_\text{U}^{(k)}  = \frac{1}{k!} \left( \frac{\mi}{2\pi} \right)^k \int_M \operatorname{Tr}(\rho \underbrace{\mathcal{F}_{\mathrm{U}} \wedge \cdots \wedge \mathcal{F}_{\mathrm{U}}}_{k}),
\end{equation}
raising the natural question of its relation to the topological invariant $n^{(k)}$ defined in Eq.\eqref{nk}.

\section{First thermal Uhlmann-Chern number}\label{fUCnum}
We begin by considering the first thermal Uhlmann-Chern number, defined for quantum systems with a two-dimensional parameter space. In particular, we investigate two distinct classes of models: two-level systems and systems with an infinite number of energy levels.

\subsection{Two-level System}\label{twolev}
\subsubsection{Berry Connection and Berry Curvature}

A generic two-level system is described by the Hamiltonian $H = R_i \sigma_i$, where $\sigma_i$ (with $i = 1, 2, 3$) are the Pauli matrices, and $\mathbf{R} = (R_1, R_2, R_3)^T$ is a vector of real parameters. If the magnitude $R = |\mathbf{R}| = \sqrt{R_1^2 + R_2^2 + R_3^2}$ is held fixed, the system effectively depends on two parameters. For instance, by setting $\mathbf{R} = R (\sin \theta \cos \phi, \sin \theta \sin \phi, \cos \theta)^T$, the parameter space becomes the two-dimensional sphere $S^2$. The eigenenergies of $H$ are $E_\pm = \pm R$, with corresponding eigenvectors given by
\begin{align}
	|\pm R\rangle = \frac{1}{\sqrt{2R(R \pm R_3)}}
	\begin{pmatrix}
		R \pm R_3 \\
		\pm (R_1 + \mi R_2)
	\end{pmatrix}.
\end{align}
A direct calculation yields the Berry connection for each eigenstate is
	\begin{equation}
		\mathcal{A}_{\mathrm{B}}^{\pm} = \langle \pm R | \mathrm{d} | \pm R \rangle = -\mathrm{i} \frac{R_2 \mathrm{d} R_1 - R_1 \mathrm{d} R_2}{2 R (R \pm R_3)}. \label{eq:Berry_A}
	\end{equation}
The corresponding Berry curvature is then given by
\begin{align}
		\mathcal{F}_{\mathrm{B}}^{\pm} &= \pm \frac{\mathrm{i}}{4} \epsilon_{ijk} \hat{R}_i \mathrm{d} \hat{R}_j \wedge \mathrm{d} \hat{R}_k, \label{eq:Berry_F_simplified}
	\end{align}
where $\hat{\mathbf{R}}=\mathbf{R}/R$ with its components being $\hat{\mathbf{R}}=(\hat{R}_1, \hat{R}_2, \hat{R}_3)$.

\subsubsection{Uhlmann Connection and Uhlmann Curvature}\label{twolevUC}
For a two-level system, the density matrix takes the form $\rho = \frac{\me^{-\beta \mathbf{R} \cdot \boldsymbol{\sigma}}}{Z}$, whose eigenvalues are
$\lambda_\pm=\frac{1}{2}\left[1\mp\tanh(\beta R)\right]$.
Introducing the projectors onto the eigenstates $P_\pm=|\pm R\rangle\langle \pm R|=\frac{1}{2}(1\pm \frac{H}{R})$, the Uhlmann connection is found to be
	\begin{align}
		\mathcal{A}_{\mathrm{U}} &=-\left(\sqrt{\lambda_+}-\sqrt{\lambda_-}\right)^2\left(P_+\dif P_-+P_-\dif P_+\right)\notag\\
&= \frac{1}{2} C (\hat{\mathbf{R}} \cdot \boldsymbol{\sigma}) (\mathrm{d} \hat{\mathbf{R}} \cdot \boldsymbol{\sigma}), \label{eq:Uhlmann_A}
	\end{align}
	where $C = 1 - \text{sech} (\beta R)$. To compute the Uhlmann curvature, we first evaluate the exterior derivative
	\begin{align}
		\mathrm{d} \mathcal{A}_{\mathrm{U}} = \frac{\mathrm{d} C \wedge (\hat{\mathbf{R}} \cdot \boldsymbol{\sigma}) (\mathrm{d} \hat{\mathbf{R}} \cdot \boldsymbol{\sigma}) + C (\mathrm{d} \hat{\mathbf{R}} \cdot \boldsymbol{\sigma}) \wedge (\mathrm{d} \hat{\mathbf{R}} \cdot \boldsymbol{\sigma})}{2}. \label{eq:Uhlmann_dA}
	\end{align}
Including the term $\mathcal{A}_{\mathrm{U}} \wedge \mathcal{A}_{\mathrm{U}}$, a straightforward albeit lengthy calculation yields the Uhlmann curvature:
\begin{align}
	\mathcal{F}_{\mathrm{U}} &= \frac{\mi\beta}{2} \text{sech}(\beta R) \tanh(\beta R) \dif R \wedge \epsilon_{abc} \hat{R}_a \dif \hat{R}_b \sigma_c \nonumber \\
& + \frac{\mi C}{2} \epsilon_{abc} \dif \hat{R}_a \wedge \dif \hat{R}_b \sigma_c -\frac{\mi C^2}{4} \epsilon_{abc} \hat{R}_a \dif \hat{R}_b \wedge \dif \hat{R}_c \hat{\mathbf{R}} \cdot \boldsymbol{\sigma}.
\end{align}
In the zero-temperature limit, $\lim_{\beta \to \infty} \beta \mathrm{sech}(\beta R) = 0$ and $\lim_{\beta \to \infty} C = 1$, hence
\begin{align}\label{twolevel}
	\text{Tr}(\rho \mathcal{F}_{\mathrm{U}}) &= -\frac{\mi}{4} \tanh^3 (\beta R) \epsilon_{abc} \hat{R}_a \dif \hat{R}_b \wedge \dif \hat{R}_c\notag\\
&\overset{T\to 0}{=}\mathcal{F}_{\mathrm{B}}^-,
\end{align}
where the last line follows by comparing with Eq.(\ref{eq:Berry_F_simplified}). Consequently, Eqs.(\ref{eq:Thermal_Ch}) and (\ref{n1}) implies that
\begin{equation}\label{twolevcor}
\lim_{T\to 0}n^{(1)}_\text{U}=n^{(1)}.
\end{equation}
In other words, the non-topological thermal Uhlmann-Chern number in this two-level model smoothly reduces to the Chern number associated with the ground state as $T\to0$. In the infinite-temperature limit, $  \rho \to \frac{1}{2}1_2  $ (where $  1_2  $ is the $2\times 2$ identity matrix), so $  \text{Tr}(\rho \mathcal{F}_\text{U}) \to 0  $ and $  \lim_{T \to \infty} n_\text{U}^{(1)} = 0  $. This is consistent with the topological triviality of the Uhlmann bundle at $  T > 0  $. Thus, $  n_\text{U}^{(1)}  $ serves as a temperature-dependent quantity connecting the topological properties of pure states at zero temperature, where it recovers the Chern number, to mixed states at finite temperatures, where it is not a topological invariant. Although the Uhlmann bundle is topologically trivial at finite temperatures, $  n_\text{U}^{(1)}  $ remains non-zero, providing a quantitative measure of how topological features evolve with temperature.
\begin{figure}[h]
	\centering
	\includegraphics[width=3.35in,clip]{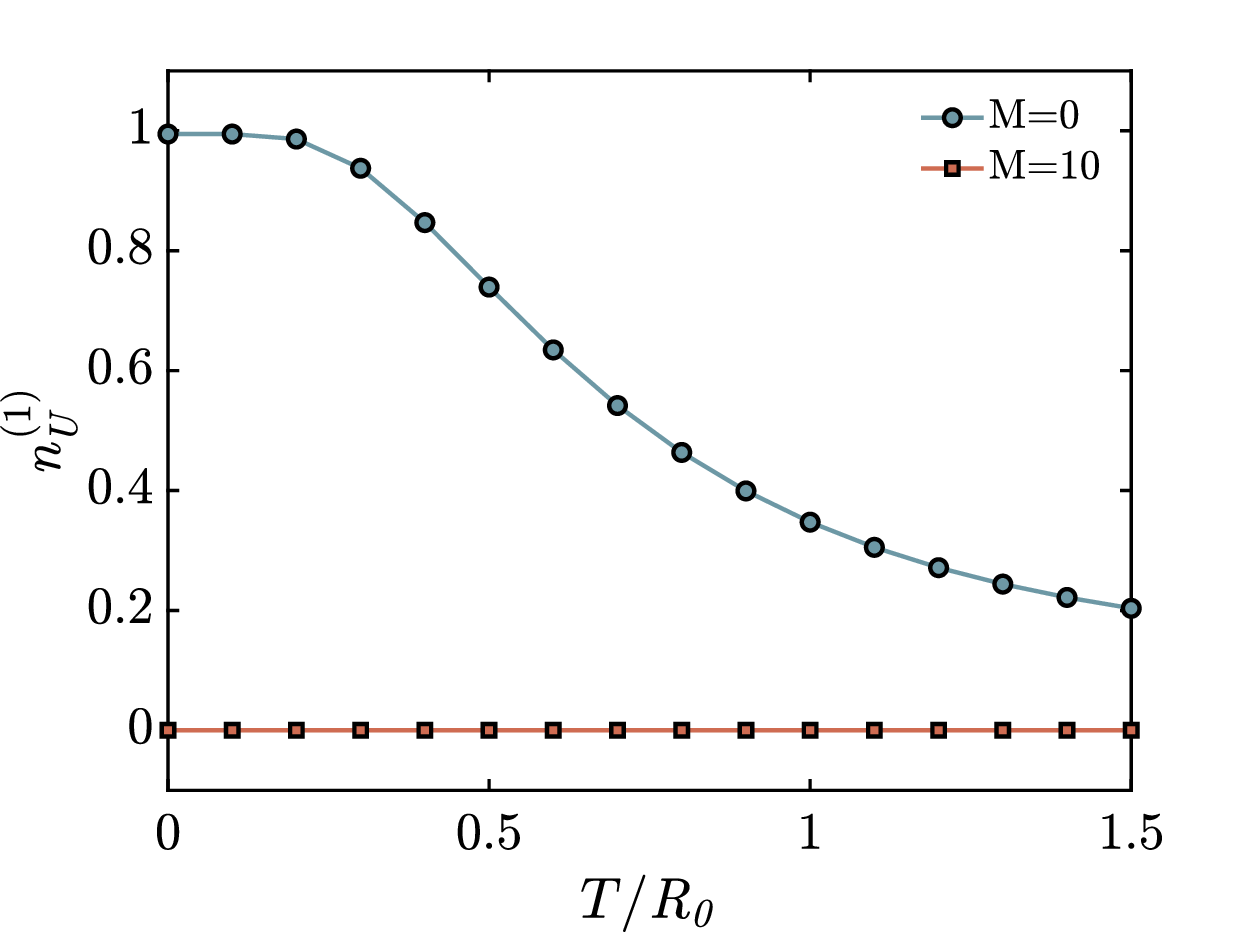}
	\caption{The first Uhlmann-Chern number of the 2D Haldane model as a function of temperature $T$ for two Semenoff mass values: $M=0$ and $M=10$, with parameters $t_1=1.0$, $t_2=0.5$, and $\phi=\pi/2$. These correspond to a topologically nontrivial phase with a nonzero first Chern number $n^{(1)}=n^{(1)}_{\mathrm{U}}(T\to 0)=1$ at the zero-temperature limit and a topologically trivial phase with $n^{(1)}=0$, respectively. The unit $R_0$ is defined as $R_0:=\Delta(M=k_x=k_y=0)=1.0$.}
	\label{Fig1}
\end{figure}
\subsection{2D Haldane Model}

To explore the correspondence in Eq.\eqref{twolevcor} as $T\to 0$, we analyze a specific two-band system: the 2D Haldane model. Introduced in 1988~\cite{PhysRevLett.61.2015}, the Haldane model serves as a paradigmatic Chern insulator, exhibiting the quantum anomalous Hall effect without requiring Landau levels. This two-dimensional lattice model features a graphene-like hexagonal structure with complex next-nearest-neighbor hopping, which breaks time-reversal symmetry while preserving translational symmetry. The Hamiltonian of the 2D Haldane model in momentum space is expressed as $H(\mathbf{k}) = \sum_{i=1}^3 R_i(\mathbf{k}) \sigma_i$, where $R_1(\mathbf{k}) = t_1 \sum_{i=1}^3 \cos(\mathbf{k}\cdot\mathbf{a}_i)$, $R_2(\mathbf{k}) = t_1 \sum_{i=1}^3 \sin(\mathbf{k}\cdot\mathbf{a}_i)$ and $R_3(\mathbf{k}) = M - 2t_2 \sin\phi \sum_{i=1}^3 \sin(\mathbf{k}\cdot\mathbf{b}_i)$. Here, $t_1$ denotes the nearest-neighbor hopping amplitude, $t_2$ represents the next-nearest-neighbor hopping, $\phi$ is the Haldane phase~\cite{PhysRevLett.61.2015, bernevig2013topological}, and $M$ is the Semenoff mass. The lattice vectors $\mathbf{a}_i$ are defined as: $\mathbf{a}_1=(\sqrt{3}, 0)^{\text{T}}$, $\mathbf{a}_2=(-\frac{\sqrt{3}}{2}, -\frac{3}{2})^{\text{T}}$, and $\mathbf{a}_3=(-\frac{\sqrt{3}}{2}, \frac{3}{2})^{\text{T}}$. The next-nearest-neighbor vectors $\mathbf{b}_i$ are given by: $\mathbf{b}_1=\mathbf{a}_2 -\mathbf{a}_3$, $\mathbf{b}_2=\mathbf{a}_1 - \mathbf{a}_2$, and $\mathbf{b}_3=\mathbf{a}_3-\mathbf{a}_1$.

\begin{figure}[ht]
	\centering
	\includegraphics[width=3.2in,clip]{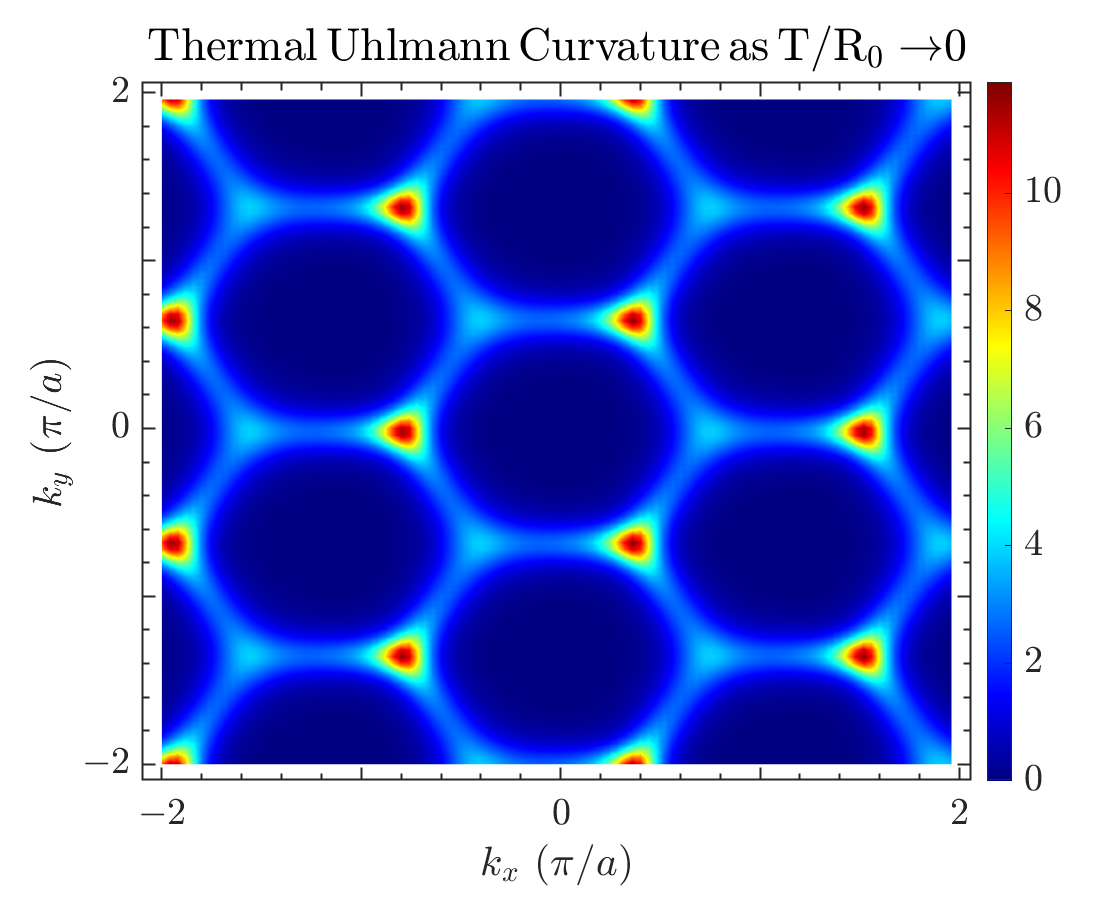}\\
	\includegraphics[width=3.2in,clip]{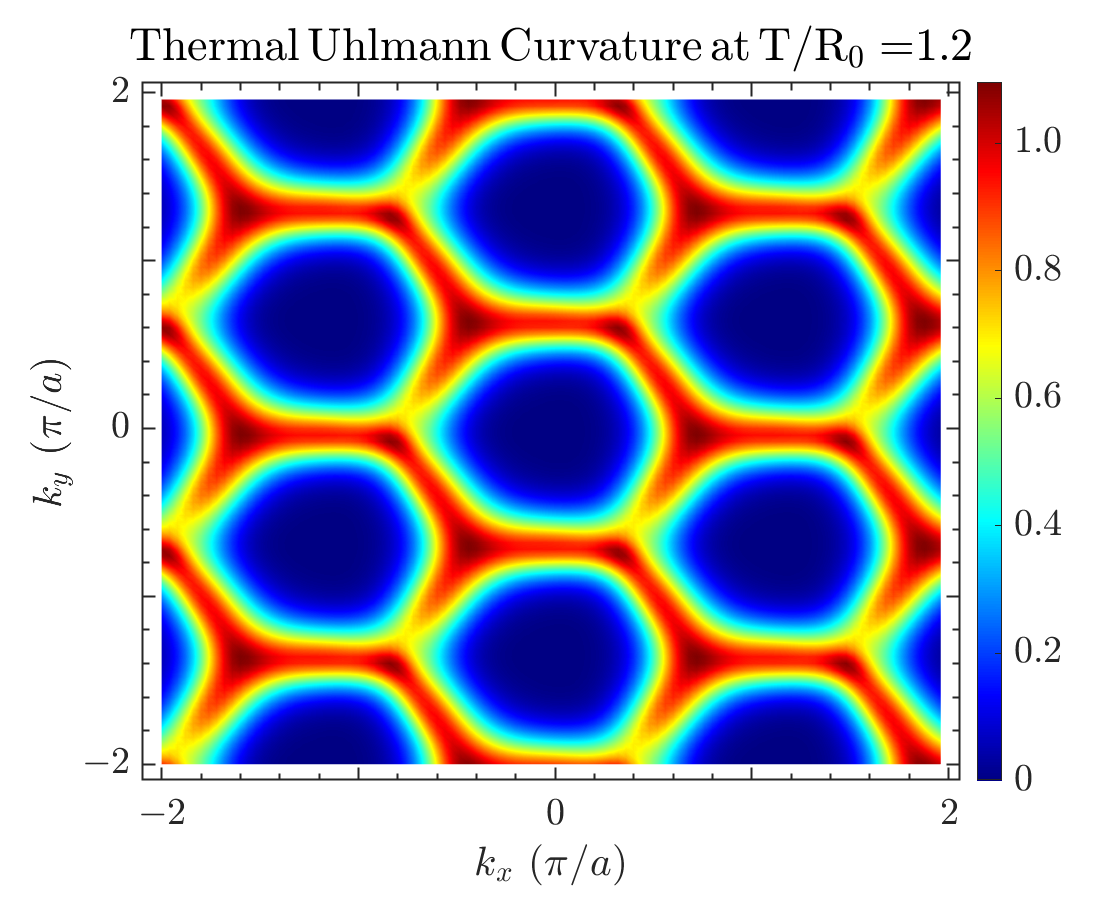}
	\caption{Comparison of the thermal Uhlmann curvature in the ground state at temperatures $T/R_0=0$ and $T/R_0=1.2$, with parameters $t_1=1$, $t_2=0.14$, $\phi=\pi/2$, and $M=-0.2$. At zero temperature, the thermal Uhlmann curvature converges to the Berry curvature~\cite{vanderbilt2018berry,PhysRevB.89.235424}, whereas at finite temperature, the geometric information encoded in the Berry curvature is progressively diminished. The unit $R_0$ is consistent with Fig.~\ref{Fig1}.}
	\label{Fig2}
\end{figure}

The thermal Uhlmann curvature for this model is given by
\begin{align}\label{Cur2DHaldane}
	\text{Tr}(\rho \mathcal{F}_{\mathrm{U}}) &= -\frac{\mi}{4} \tanh^3 \left(\frac{\beta \Delta_k}{2}\right) \epsilon_{abc} \hat{R}_a \dif \hat{R}_b \wedge \dif \hat{R}_c,
\end{align}
where $\Delta_k=2\sqrt{R^2_1(\mathbf{k})+R^2_2(\mathbf{k})+R^2_3(\mathbf{k})}$ is the energy gap function, and $\hat{R}_i \equiv 2R_i/\Delta_k$. Consequently, the first Uhlmann-Chern number is
\begin{align}\label{UCn2DHaldane}
	n_\mathrm{U}^{(1)} &= \frac{\mi}{2\pi}\int \text{Tr}(\rho \mathcal{F}_{\mathrm{U}}) \notag \\
	&= \frac{1}{8\pi} \int \tanh^3 \left(\frac{\beta \Delta_k}{2}\right) \epsilon_{abc} \hat{R}_a \dif \hat{R}_b \wedge \dif \hat{R}_c.
\end{align}
As depicted in Fig.~\ref{Fig1}, the first-order Uhlmann-Chern number $n_\mathrm{U}^{(1)}$ is plotted as a function of temperature $T$ for two values of the Semenoff mass: $M=0$ and $M=10$. The figure reveals that $n_\mathrm{U}^{(1)}$ approaches zero at infinite temperature, irrespective of $M$. This behavior is expected, as the density matrix $\rho$ becomes proportional to the identity matrix at $T \to \infty$, causing the integrand in Eq.~\eqref{UCn2DHaldane} to vanish, resulting in $\lim_{T \rightarrow \infty} n_\mathrm{U}^{(1)} = 0$. This trend is governed by the factor $\tanh^3(\Delta_k/2T)$, which approaches zero at high temperatures. In the zero-temperature limit, $n_\mathrm{U}^{(1)}$ reduces to the conventional Chern number $n^{(1)}$, i.e., $n_\mathrm{U}^{(1)}\overset{T\to 0}{=}n^{(1)}$. For $M=0$, satisfying $|M| < 3\sqrt{3}t_2|\sin\phi|$, the system exhibits a topologically nontrivial phase with $n^{(1)}=1$, while for $M=10$, where $|M| > 3\sqrt{3}t_2|\sin\phi|$, it is topologically trivial with $n^{(1)}=0$. These numerical results align precisely with theoretical predictions. The temperature dependence of $ n_\mathrm{U}^{(1)} $ connects the maximally mixed state at high temperatures to the pure state at zero temperature, capturing the transition at finite temperatures.

To further illustrate the topological properties, we plot the thermal Uhlmann curvature across the Brillouin zone. As shown in Fig.~\ref{Fig2}, at the zero-temperature limit ($T/R_0 \to 0$), the thermal Uhlmann curvature exactly reproduces the conventional Berry curvature. However, at elevated temperatures ($T/R_0=1.2$ in our calculations), the geometric information encoded in the Berry curvature is gradually lost, reflecting the temperature dependence of the Uhlmann-Chern number, where thermal fluctuations progressively suppress the system's topological characteristics.

\subsection{Coherent State}

Previous studies of low-dimensional systems have focused on systems with finite energy levels. To explore a more complex example, we now consider an infinite-dimensional system: coherent states, which can be constructed from harmonic oscillators. The Hamiltonian of a single harmonic oscillator is given by $H = \hbar\omega\left(a^\dagger a + \frac{1}{2}\right)$, where $a$ and $a^\dagger$ are the annihilation and creation operators satisfying the commutation relation $[a, a^\dagger] = 1$. The bosonic harmonic oscillator provides an infinite-dimensional example through the translation operator $D(z) \equiv \me^{z a^\dagger - \bar{z} a}$. One can verify the following relation:
\begin{align}\label{DzdDz}
	D^{\dagger}(z) \dif D(z) = \left(a^{\dagger} + \frac{1}{2}\bar{z}\right) \dif z - \left(a + \frac{1}{2}z\right) \dif \bar{z}.
\end{align}
The state $|z\rangle$ is the ground state of the translated Hamiltonian $H(z) = D(z) H D^{\dagger}(z)$. The excited states are obtained similarly: $|n,z\rangle = D(z)|n\rangle$ for $n = 0, 1, 2, \dots$, where $|n\rangle$ are the eigenstates of $H$.

\subsubsection{Berry Connection and Berry Curvature}\label{bcsBC}
The parameter space of this model is the complex $z$-plane. It is noteworthy that, despite the infinite number of energy levels in the coherent state model, its parameter space remains two-dimensional. Consequently, only the first Chern number is well-defined and physically meaningful. Using Eq.~\eqref{DzdDz}, the Berry connection and Berry curvature are given by:
\begin{align}
	\mathcal{A}^n_{\mathrm{B}} &= \langle n | D^{\dagger}(z) \dif D(z) |n\rangle = \frac{1}{2} (\bar{z} \dif z - z \dif \bar{z}), \notag \\
	\mathcal{F}^n_{\mathrm{B}} &= \dif \mathcal{A}^n_{\mathrm{B}} = \dif \bar{z} \wedge \dif z, \label{coherentBcurvature}
\end{align}
where the superscript $n$ denotes the $n$-th energy level.

\subsubsection{Uhlmann Connection and Uhlmann Curvature}\label{bcsUC}
For a canonical ensemble at temperature $T$, the density matrix of the coherent state system is:
\begin{align}
	\rho(z) = \frac{1}{Z} \me^{-\beta H(z)} = D(z) \rho(0) D^{\dagger}(z),
\end{align}
where $\rho(0) = \frac{1}{Z} \me^{-\beta H}$. Using Eq.~\eqref{ucon}, the Uhlmann connection is:
\begin{align}
	\mathcal{A}_\mathrm{U} &= -\sum_{n \neq m} \frac{(\sqrt{\lambda_n} - \sqrt{\lambda_m})^2}{\lambda_n + \lambda_m} |n,z\rangle \langle n,z| \dif |m,z\rangle \langle m,z| \notag \\
	&= -\sum_{n \neq m} \chi_{nm} D(z) |n\rangle \langle n| D^{\dagger}(z) \dif D(z) |m\rangle \langle m| D^{\dagger}(z),
\end{align}
where $\chi_{nm} = \frac{(\me^{-\frac{n}{2}\beta \hbar \omega} - \me^{-\frac{m}{2}\beta \hbar \omega})^2}{\me^{-n \beta \hbar \omega} + \me^{-m \beta \hbar \omega}}$. To compute the Uhlmann curvature $\mathcal{F}_\mathrm{U} = \dif \mathcal{A}_\mathrm{U} + \mathcal{A}_\mathrm{U} \wedge \mathcal{A}_\mathrm{U}$, we introduce $f(\beta) = 1 - \text{sech} \frac{\beta \hbar \omega}{2}$. The two terms of the Uhlmann curvature are:
\begin{align}
	\dif \mathcal{A}_\mathrm{U} &= -2 f(\beta) \dif z \wedge \dif \bar{z} \, 1_{\infty}, \notag \\
	\mathcal{A}_\mathrm{U} \wedge \mathcal{A}_\mathrm{U} &= f(\beta)^2 \dif z \wedge \dif \bar{z} \, 1_{\infty}.
\end{align}
Thus,
\begin{align}
	\mathcal{F}_\mathrm{U} = -\tanh^2 \frac{\beta \hbar \omega}{2} \dif z \wedge \dif \bar{z} \, 1_{\infty}.
\end{align}
Here, we explicitly denote the infinite-dimensional identity matrix $1_{\infty}$. The Uhlmann curvature satisfies $\mathcal{F}_\mathrm{U}^\dagger = -\mathcal{F}_\mathrm{U}$ and is proportional to $1_{\infty}$, rendering $\text{Tr}(\mathcal{F}_\mathrm{U})$ ill-defined. However, $\text{Tr}(\rho(z) \mathcal{F}_\mathrm{U})$ is well-defined:
\begin{align}
	\text{Tr}(\rho(z) \mathcal{F}_\mathrm{U}) &= -\tanh^2 \frac{\beta \hbar \omega}{2} \dif z \wedge \dif \bar{z} \, \text{Tr}(\rho(z)) \notag \\
	&= -\tanh^2 \frac{\beta \hbar \omega}{2} \dif z \wedge \dif \bar{z}.
\end{align}
In the zero-temperature limit, $\tanh^2 \frac{\beta \hbar \omega}{2} \to 1$, so $\text{Tr}(\rho(z) \mathcal{F}_\mathrm{U}) \to -\dif z \wedge \dif \bar{z}$. Comparing with the Berry curvature in Eq.~\eqref{coherentBcurvature}, we find:
\begin{align}
	\lim_{T \to 0} \text{Tr}(\rho(z) \mathcal{F}_\mathrm{U}) = \mathcal{F}^0_{\mathrm{B}},
\end{align}
where $\mathcal{F}^0_{\mathrm{B}}$ is the Berry curvature of the ground state. This result is consistent with the finite-dimensional case.

\section{Proof of the Correspondence between $n_\text{U}^{(1)}$ and $n^{(1)}$ as $T\to 0$}\label{prooffiniteD}
In the preceding sections, we have established that $\operatorname{Tr}(\rho \mathcal{F}_{\mathrm{U}}) \to \mathcal{F}_{\mathrm{B}}$ for both the two-level model and the coherent state system, suggesting a general mathematical proof. Here, we provide a universal proof for discrete quantum systems. We start with:
\begin{align}
	&\text{Tr}(\rho \mathcal{F}_{\mathrm{U}}) = \sum_i \lambda_i \langle i | \mathcal{F}_\mathrm{U} | i \rangle \notag \\
	=& \sum_i \lambda_i \left[ \dif (\mathcal{A}_{\mathrm{U}})_{ii} + \sum_k (\mathcal{A}_{\mathrm{U}})_{ik} \wedge (\mathcal{A}_\mathrm{U})_{ki} \right].
\end{align}
We examine the contributions of both terms separately. Define $C_{ij} = \frac{(\sqrt{\lambda_i} - \sqrt{\lambda_j})^2}{\lambda_i + \lambda_j}$, which satisfies $C_{ii} = 0$ and $C_{ij} = C_{ji}$. For $\mathcal{A}_{\mathrm{U}} = -\sum_{j \neq k} C_{jk} |j\rangle \langle j| \dif |k\rangle \langle k|$, its exterior derivative is:
\begin{align}\label{dAu}
	\dif \mathcal{A}_{\mathrm{U}} &= -\sum_{j \neq k} \dif C_{jk} |j\rangle \langle j| \dif k\rangle \langle k| - \sum_{j \neq k} C_{jk} \big( |\dif j\rangle \wedge \langle j| \dif k\rangle \langle k| \notag \\
	&\quad + |j\rangle \langle \dif j| \wedge |\dif k\rangle \langle k| + |j\rangle \langle \dif k| \wedge \langle j| \dif k\rangle \big).
\end{align}
Here, $|\dif k\rangle \equiv \dif |k\rangle$ and $\langle \dif k| \equiv \dif \langle k|$. When computing $\text{Tr}(\rho \dif \mathcal{A}_{\mathrm{U}})$, the first term in Eq.~\eqref{dAu} gives:
\begin{align}
	&-\sum_i \lambda_i \sum_{j \neq k} \dif C_{jk} \langle i | j \rangle \langle j | \dif  k \rangle \langle k | i \rangle\notag\\ =& -\sum_i \lambda_i \sum_{j \neq k} \dif C_{jk} \langle j | \dif  k \rangle \delta_{ij} \delta_{ki}.
\end{align}
This term vanishes since $j \neq k$ precludes $i = j = k$. Similarly, the third term in Eq.~\eqref{dAu} yields zero under the trace. Thus:
\begin{align}
	&\text{Tr}(\rho \dif \mathcal{A}_{\mathrm{U}}) \notag\\=& -\sum_i \lambda_i \sum_{j \neq k} \big( C_{ji} \langle i | \dif  j \rangle \wedge \langle j | \dif i \rangle + C_{ik} \langle \dif  k | i \rangle \wedge \langle i | \dif  k \rangle \big) \notag \\
	=& -2 \sum_{i \neq k} \lambda_i \frac{(\sqrt{\lambda_i} - \sqrt{\lambda_k})^2}{\lambda_i + \lambda_k} \langle i | \dif  k \rangle \wedge \langle k | \dif i \rangle.
\end{align}
The second term of $\text{Tr}(\rho \mathcal{F}_{\mathrm{U}})$ is:
\begin{align}
	\text{Tr}(\rho \mathcal{A}_{\mathrm{U}} \wedge \mathcal{A}_{\mathrm{U}}) = \sum_{k \neq i} \lambda_i \frac{(\sqrt{\lambda_k} - \sqrt{\lambda_i})^4}{(\lambda_i + \lambda_k)^2} \langle i | \dif | k \rangle \wedge \langle k | \dif | i \rangle.
\end{align}
Combining both terms, we obtain:
\begin{align}
	\text{Tr}(\rho \mathcal{F}_{\mathrm{U}}) = \sum_{k \neq i} \lambda_i \left[ \frac{4 \lambda_i \lambda_k}{(\lambda_i + \lambda_k)^2} - 1 \right] \langle i | \dif | k \rangle \wedge \langle k | \dif | i \rangle.
\end{align}
In the zero-temperature limit, where $E_0 < E_1 < \dots$, only the ground state occupation probability $\lambda_0$ approaches 1, while $\lambda_{k>0} \to 0$. Thus, the dominant contribution comes from $i = 0$:
\begin{align}
	\text{Tr}(\rho \mathcal{F}_{\mathrm{U}}) \to \lambda_0 \langle 0 | \mathcal{F}_{\mathrm{U}} | 0 \rangle.
\end{align}
At zero temperature, with $\lambda_0 \to 1$ and $\lambda_{k>0} \to 0$, we have:
\begin{align}
	\lim_{T \to 0} \frac{4 \lambda_i \lambda_{k>0}}{(\lambda_i + \lambda_{k>0})^2} = 0.
\end{align}
Consequently:
\begin{align}
	\text{Tr}(\rho \mathcal{F}_{\mathrm{U}}) \to -\sum_{k \neq 0} \langle 0 | \dif | k \rangle \wedge \langle k | \dif | 0 \rangle.
\end{align}
Meanwhile, the Berry curvature is expressed as:
\begin{align}
	\mathcal{F}_{\mathrm{B}} &= \mathrm{d} \langle 0 | \wedge \mathrm{d} | 0 \rangle \notag \\
	&= \sum_{mn} (\mathrm{d} \langle 0 | ) | m \rangle \langle m | \wedge | n \rangle \langle n | \mathrm{d} | 0 \rangle \notag \\
	&= -\sum_{k \neq 0} \langle 0 | \mathrm{d} | k \rangle \wedge \langle k | \mathrm{d} | 0 \rangle. \label{eq:Appendix_F_B}
\end{align}
Thus, we have proven that the Uhlmann curvature reduces to the Berry curvature in the zero-temperature limit:
\begin{equation}
	\operatorname{Tr}(\rho \mathcal{F}_{\mathrm{U}}) \to \mathcal{F}_{\mathrm{B}}. \label{eq:Appendix_Tr4}
\end{equation}
This demonstrates that the Uhlmann curvature converges to the Berry curvature at zero temperature, which further implies $n_\text{U}^{(1)}\to n^{(1)}$.

\section{Second Thermal Uhlmann-Chern Number of a Four-Band Model}\label{secUCnum}
Consider a quantum system with a Hamiltonian expressed as $H = \sum_{i=1}^{5} R_i \Gamma_i$, where the Gamma matrices are $\Gamma_i = \sigma_1 \otimes \sigma_i$ for $i = 1, 2, 3$; $\Gamma_4 = \sigma_2 \otimes 1$, and $\Gamma_5 = \sigma_3 \otimes 1$. These satisfy the anticommutation relations $\{\Gamma_i, \Gamma_j\} = 2 \delta_{ij}$. The four energy levels are:
\begin{align}
	|\psi_{a,c}\rangle &= \frac{1}{\sqrt{2R(R \mp R_5)}}
	\begin{pmatrix}
		-R_3 + \mi R_4 \\
		-R_1 - \mi R_2 \\
		R_5 \mp R \\
		0
	\end{pmatrix}, \notag \\
	|\psi_{b,d}\rangle &= \frac{1}{\sqrt{2R(R \mp R_5)}}
	\begin{pmatrix}
		-R_1 + \mi R_2 \\
		R_3 + \mi R_4 \\
		0 \\
		R_5 \mp R
	\end{pmatrix},
\end{align}
where $R = \sqrt{\sum_{i=1}^{5} R_i^2}$. These four levels are doubly degenerate and satisfy:
\begin{equation}
	H |\psi_{a,b}\rangle = R |\psi_{a,b}\rangle, \quad H |\psi_{c,d}\rangle = -R |\psi_{c,d}\rangle.
\end{equation}
The Hamiltonian depends on the parameters $R = (R_1, R_2, R_3, R_4, R_5)^T$. The Boltzmann weights of the two degenerate subspaces are:
\begin{align}
	p_{1,2} = \frac{2 \me^{\pm R/T}}{Z},
\end{align}
with the partition function $Z = 4 \cosh(R/T)$. The projection operators for the subspaces and the density matrix are given by:
\begin{align}
	P_1 &= |\psi_a\rangle \langle \psi_a| + |\psi_b\rangle \langle \psi_b| = \frac{1}{2} \left( 1_4 + \hat{R}_i \Gamma_i \right), \\
	P_2 &= |\psi_c\rangle \langle \psi_c| + |\psi_d\rangle \langle \psi_d| = \frac{1}{2} \left( 1_4 - \hat{R}_i \Gamma_i \right), \\
	\rho &= \frac{p_i}{2} P_i = \frac{1}{4} \left[ 1_4 + \tanh \left( \frac{R}{T} \right) \hat{R}_i \Gamma_i \right],
\end{align}
where $\hat{R}_i = R_i / R$.
The Uhlmann connection $\mathcal{A}_\mathrm{U}$ is found to be:
\begin{align}
	\mathcal{A}_\mathrm{U} &= -f(R) ( P_1 \dif P_2 + P_2 \dif P_1 ) \notag \\
	&= -\frac{\mi}{4} f(R) ( \hat{R}_a \dif \hat{R}_b - \hat{R}_b \dif \hat{R}_a ) \Gamma_{ab}.
\end{align}
Here, $f(R) = 1 - \text{sech}(R/T)$, and $\Gamma_{ab} = \mi [\Gamma_a, \Gamma_b]/2$, satisfying $\Gamma_{ab}^2 = 1_4$. We have used the identity $\hat{R}_a \dif \hat{R}_a = 0$. The Uhlmann curvature is defined as $\mathcal{F}_\mathrm{U} = \dif \mathcal{A}_\mathrm{U} + \mathcal{A}_\mathrm{U} \wedge \mathcal{A}_\mathrm{U}$.
The second thermal Uhlmann-Chern number in this model is:
\begin{align}\label{seUC}
	n_\text{U}^{(2)} &= -\frac{1}{8 \pi^2} \int \text{tr}(\rho \mathcal{F}_\mathrm{U} \wedge \mathcal{F}_\mathrm{U}) \notag \\
	&= \frac{3}{16 \pi^2 \cdot 4!} \int \epsilon_{abcde} \hat{R}_a \dif \hat{R}_b \dif \hat{R}_c \dif \hat{R}_d \dif \hat{R}_e \tanh^5 \left( \frac{R}{T} \right),
\end{align}
where $\epsilon_{abcde}$ is the Levi-Civita symbol, and $4!$ arises from distinct permutations of the four exterior differential terms.

\begin{figure}[t]
	\centering
	\includegraphics[width=3.35in,clip]{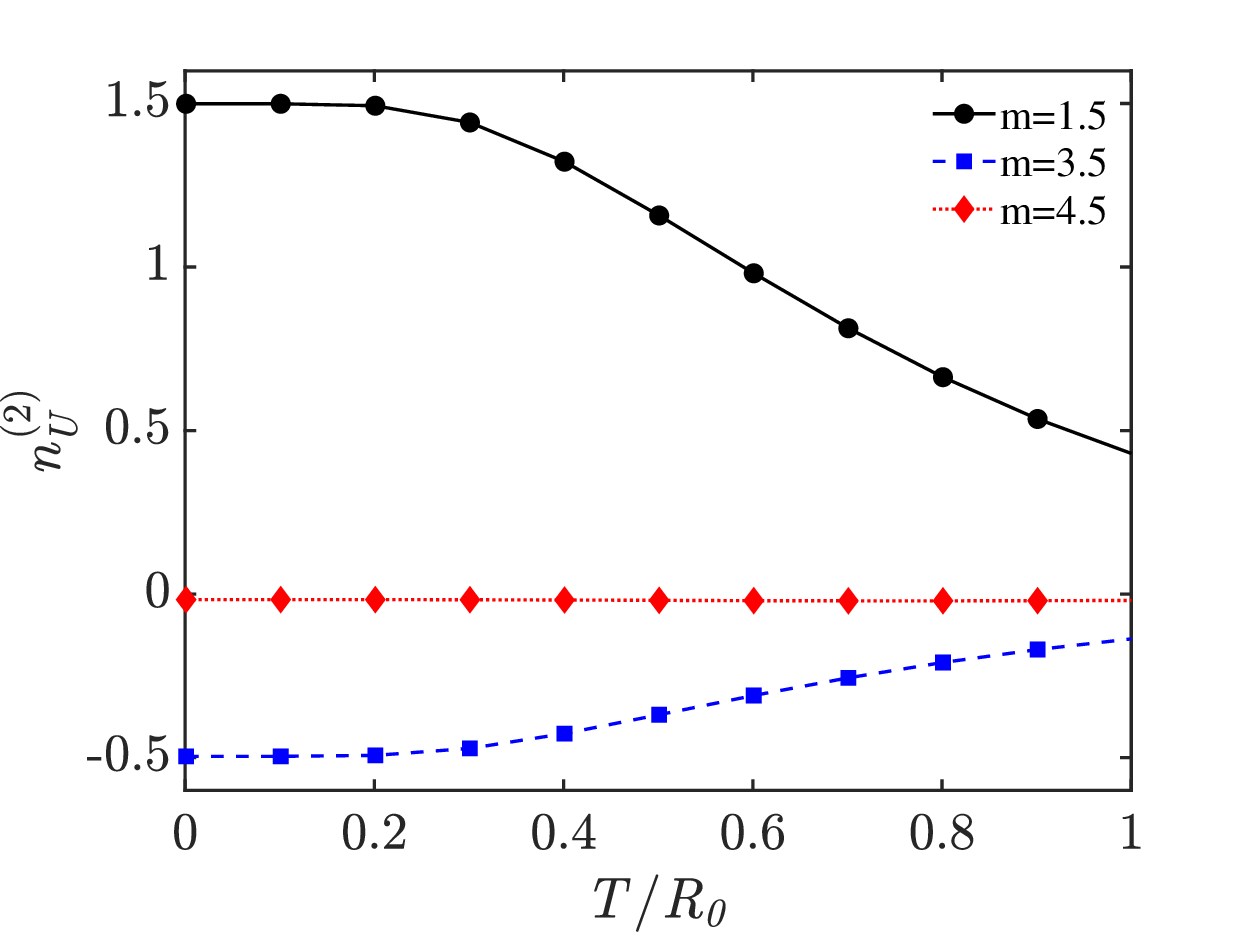}
	\caption{The second Uhlmann-Chern number as a function of temperature $T$ for three different $m$-values: $m=1.5$, $3.5$, and $4.5$, corresponding to distinct topological phases. The unit $R_0$ is defined as $R_0 := R(k_i = \pi/4, m = -3) = 1.0$, where $k_i = k_x, k_y, k_z, k_w$.}
	\label{Fig3}
\end{figure}

We consider a specific four-dimensional tight-binding model designed to study 4D quantum Hall effects. The parameters $R_i$ depend on a four-dimensional parameter space ($k_x, k_y, k_z, k_w$), with $R_1 = \cos 2k_x$, $R_2 = \cos 2k_y$, $R_3 = \cos 2k_z$, $R_4 = \cos 2k_w$, and $R_5 = m + \sin 2k_x + \sin 2k_y + \sin 2k_z + \sin 2k_w$. The second thermal Uhlmann-Chern number becomes:
\begin{align}
	 n_\text{U}^{(2)}&= \frac{3}{\pi^2} \int \frac{\dif k_x\, \dif k_y\,  \dif k_z\,  \dif k_w}{R^5}
	  ( \sin 2 k_x \sin 2  k_y \sin 2  k_z\notag\\
	  & +\sin 2  k_x \sin 2  k_z \sin 2  k_w
	+\sin 2  k_y \sin 2  k_z \sin 2  k_w \notag\\
	&+ m \sin 2  k_x \sin 2  k_y \sin 2  k_z \sin 2  k_w )
	\tanh^5 \left(\frac{R}{T}\right).
\end{align}
Similar to the first Uhlmann-Chern number, numerical calculations yield the temperature-dependent second Uhlmann-Chern number $n_\text{U}^{(2)}$. As shown in Fig.~\ref{Fig3}, $n_\text{U}^{(2)}$ approaches $n^{(2)}/2$ in the zero-temperature limit, where the factor $D=2$ represents the degeneracy of the ground state in this four-band model. This result suggests a general relationship between the $n$-th Uhlmann-Chern number and the Chern number at $T/R_0 \to 0$, which we will prove in the next section. The second-order Chern number for an equivalent 4D model is given in~\cite{PhysRevB.78.195424, PhysRevB.97.235141}:
\begin{align}
	n^{(2)} =
	\begin{cases}
		3, & \text{for } 0 < m < 2; \\
		-3, & \text{for } -2 < m < 0; \\
		-1, & \text{for } 2 < m < 4; \\
		1, & \text{for } -4 < m < -2; \\
		0, & \text{for } |m| > 4,
	\end{cases}
\end{align}
where different values of $m$ correspond to distinct topological phases. Clearly, $n_\text{U}^{(2)}$ reflects the topological properties of the system at zero temperatures, while the geometric information encoded in the Chern number is gradually lost as temperature increases. At high temperatures, as $T \to \infty$, the factor $\tanh^5 (R/T) \to 0$, rendering $\text{tr}(\rho \mathcal{F}_\mathrm{U} \wedge \mathcal{F}_\mathrm{U})$ proportional to $\text{tr}(\mathcal{F}_\mathrm{U} \wedge \mathcal{F}_\mathrm{U})$, which vanishes due to the triviality of the Uhlmann bundle.

\section{Proof of the Correspondence between $n_\text{U}^{(k)}$ and $n^{(k)}$ as $T\to 0$}\label{UWCcurv}
In Section~\ref{prooffiniteD}, we established the relationship in Eq.~\eqref{eq:Appendix_Tr4} between the Uhlmann curvature and Berry curvature in the zero-temperature limit, which can be expressed as
\begin{equation}\label{eqB1}
	\lim_{T \rightarrow 0}  \langle 0| \mathcal{F}_{\mathrm{U}} |0 \rangle  = \mathcal{F}_{\mathrm{B}},
\end{equation}
or equivalently written in a concise form
\begin{align}\label{eqB2}
	\lim_{T \rightarrow 0} \operatorname{Tr} \left( \rho \mathcal{F}_{\mathrm{U}} \right)=\lim_{T \rightarrow 0}  \langle W| \mathcal{F}_{\mathrm{U}} |W\rangle = \mathcal{F}_{\mathrm{B}} ,
\end{align}
by using the purified state $|W\rangle = \sum_{i} \sqrt{\lambda_i}|i\rangle \otimes U^T|i\rangle$. In the first equation, we have applied
\begin{align}
	\langle W| \mathcal{F}_{\mathrm{U}} |W\rangle &=\sum_{i,j} \sqrt{\lambda_i \lambda_j}  \langle j| \mathcal{F}_{\mathrm{U}} |i \rangle  \otimes  \langle j|i \rangle \notag\\
	&= 	\operatorname{Tr} \left( \rho \mathcal{F}_{\mathrm{U}} \right).
\end{align}
Here, we assume that $\mathcal{F}_{\mathrm{U}}$ acts only on the system’s space and not on the ancilla space, expressed as $\mathcal{F}_{\mathrm{U}} |W\rangle := \left( \mathcal{F}_{\mathrm{U}} \otimes \hat{I} \right) |W\rangle$, where $\hat{I}$ is the identity operator in the ancilla space.

Results in Section~\ref{secUCnum} suggest a general correspondence in systems with a $D$-fold degenerate ground state. To prove this, consider a quantum system with a $D$-fold degenerate ground state subspace, with the density matrix
\begin{equation}
	\rho = \left[ \sum_{a=1}^{D} \lambda_0 |\psi_a\rangle \langle \psi_a| + \sum_{\mu \notin \{1, \ldots, D\}} \lambda_\mu |\psi_\mu\rangle \langle \psi_\mu| \right]
\end{equation}
In the zero-temperature limit, the ground state occupation probabilities satisfy $\lambda_a \to \frac{1}{D}$ ($a = 1, \dots, D$), while the excited state probabilities satisfy $\lambda_\mu \to 0$ ($\mu > D$). The Uhlmann connection $\mathcal{A}_{\mathrm{U}}$ in the zero-temperature limit approximates to
\begin{align}\label{AU}
	\mathcal{A}_{\mathrm{U}} = -\sum_{a,\mu} \left(|\psi_a\rangle\langle\psi_a|\dif \psi_\mu\rangle\langle\psi_\mu| + |\psi_\mu\rangle\langle\psi_\mu|\dif \psi_a\rangle\langle\psi_a|\right),\notag\\
\end{align}
since the terms within the ground state subspace vanish as $C_{ab} \to 0$, while the cross terms between the ground and excited states satisfy $C_{a\mu} = C_{\mu a} \to 1$. 
In the third term of Eq. (\ref{AU}), the coefficient $ C_{\mu\nu} $ in the zero-temperature limit requires case-by-case analysis. If the quantum numbers $ \mu $ and $ \nu $ belong to the same degenerate excited-state subspace (if such a subspace exists), i.e., when $ \lambda_\mu $ and $ \lambda_\nu $ correspond to the same energy level, then $ C_{\mu\nu} = 0 $. Conversely, when $ \lambda_\mu $ and $ \lambda_\nu $ correspond to different energy levels, it can be proven that $ C_{\mu\nu} = 1 $. Here, our goal is to generalize Eq. (\ref{eqB1}) by replacing the left-hand-side by $ \langle \psi_a | \mathcal{F}_\text{U} | \psi_b \rangle $.
It is noteworthy that the matrix structure of the third term in Eq. (\ref{AU}) is of the form $ |\psi_\mu\rangle \langle \psi_\nu| $. When computing $ \mathcal{F}_\text{U} $, the contribution from this term always involves at least one of $ |\psi_\mu\rangle $ or $ \langle \psi_\nu| $. Consequently, when calculating the matrix elements of $ \mathcal{F}_\text{U} $ within the ground-state subspace, the contribution of this term is necessarily zero. Therefore, this kind of terms can be safely neglected in subsequent calculations.
Introducing the ground-state projection operator $ P = \sum_a |\psi_a\rangle \langle \psi_a| $ and the excited-state projection operator $ Q = \sum_\mu |\psi_\mu\rangle \langle \psi_\mu| $, it is evident that $ P + Q = 1 $. Using the orthogonality condition $ \langle \psi_a | \psi_\mu \rangle = 0 $, it follows that in the zero-temperature limit (and hereafter), we have
\begin{align}
	\mathcal{A}_{\mathrm{U}} = -P \dif Q - Q \dif P,
\end{align}
where any term related to $ |\psi_\mu\rangle $ or $ \langle \psi_\nu| $ has been neglected as stated before. 
To compute $\mathcal{F}_{\mathrm{U}}$, we note
\begin{align}
	\dif  \mathcal{A}_{\mathrm{U}} = - \dif P \land  \dif Q -  \dif Q \land  \dif P = 2 \dif P \land  \dif P,
\end{align}
which can be obtained by using the property $P^2 = P$. Differentiating $P^2 = P$ yields $(\dif P) P + P \dif P = \dif P$, implying $P (\dif P) P = 0$. Additionally, we have $\mathcal{A}_{\mathrm{U}} = 2 P \dif P - \dif P$, and thereby
\begin{align}
	\mathcal{A}_{\mathrm{U}} \land \mathcal{A}_{\mathrm{U}} &= 4P \dif P \land P \dif P - 2P \dif P \land  \dif P \notag\\
	&\quad- 2 \dif P \land P \dif P +  \dif P \land  \dif P\notag\\
	&=-\dif P \land  \dif P.
\end{align}
Combining these results, we obtain
\begin{align}
	\mathcal{F}_{\mathrm{U}} =  \dif \mathcal{A}_{\mathrm{U}} + \mathcal{A}_{\mathrm{U}} \land \mathcal{A}_{\mathrm{U}} =  \dif P \land  \dif P.
\end{align}
Using $\dif P = \sum_a \left( | \dif \psi_a \rangle \langle \psi_a | + | \psi_a \rangle \langle \dif \psi_a | \right)$, we find
\begin{align}
	&\mathcal{F}_{\mathrm{U}} = \sum_{c,d} (| \dif \psi_c\rangle \land \langle\psi_c| \dif \psi_d\rangle\langle\psi_d| +| \dif \psi_c\rangle \land \langle  \dif \psi_d| \delta_{cd}\notag\\
	 +& |\psi_c\rangle \langle  \dif \psi_c|\wedge | \dif \psi_d\rangle\langle\psi_d|  + |\psi_c\rangle \langle  \dif \psi_c|\psi_d\rangle \wedge  \langle \dif \psi_d|).
\end{align}
In the ground state subspace, the matrix elements of $\mathcal{F}_{\mathrm{U}}$ in the zero-temperature limit are
\begin{align}
	\mathcal{F}_{\mathrm{U},ab} &= \langle\psi_a|\mathcal{F}_{\mathrm{U}}|\psi_b\rangle = \langle  \dif \psi_a| \land | \dif \psi_b\rangle \notag\\
	&+\quad  \sum_c \langle\psi_a| \dif \psi_c\rangle \land \langle\psi_c| \dif \psi_b\rangle.
\end{align}
Using the Wilczek-Zee connection, $\mathcal{A}_{ab}^{\text{WZ}} = \langle \psi_a | \dif \psi_b \rangle$, we obtain
\begin{equation}\label{WZcurv}
	\lim_{T \to 0} \langle \psi_a | \mathcal{F}_{\mathrm{U}} | \psi_b \rangle= \mathcal{F}_{ab}^{\text{WZ}}
\end{equation}
This generalizes Eq.~\eqref{eqB1}. When the ground state degeneracy is 1, this reduces to Eq.~\eqref{eqB1}. By analogy with the $k$-th Chern number, we define the $k$-th non-Abelian Uhlmann-Chern number
\begin{equation}
	n_\text{U}^{(k)} = \frac{1}{k!} \left( \frac{\mi}{2\pi} \right)^k \int_M \operatorname{Tr}(\rho \underbrace{\mathcal{F}_{\mathrm{U}} \wedge \cdots \wedge \mathcal{F}_{\mathrm{U}}}_{k}),
\end{equation}
where $M$ is the parameter manifold. In the zero-temperature limit, $\rho \to \frac{1}{D} P$, and the trace is performed only in the ground state subspace. Using Eq.~\eqref{WZcurv}, we obtain
\begin{align}
	n_\text{U}^{(k)} &\overset{T\to 0}{=} \frac{1}{D\cdot k!} \left( \frac{\mi}{2\pi} \right)^k \int_M \mathcal{F}_{a_1 a_2}^{\text{WZ}} \wedge \mathcal{F}_{a_2 a_3}^{\text{WZ}} \wedge \cdots \wedge \mathcal{F}_{a_k a_1}^{\text{WZ}} \notag\\
	&\,\,\,= \frac{n^{(k)}}{D}.
\end{align}
Here, $n^{(k)}$ is the $k$-th Chern number.

\section{Conclusion}\label{Sec.6}
In this work, we have developed a comprehensive framework for characterizing the topological properties of mixed states at finite temperatures through the thermal Uhlmann-Chern number. By incorporating the density matrix into the Chern character, this quantity captures the geometric and topological features of mixed states, connecting the pure-state Chern number at zero temperature to the trivial state at infinite temperature. A key achievement of this study is the rigorous mathematical proof, detailed in Section~\ref{UWCcurv}, that the first and higher-order Uhlmann-Chern numbers converge to the corresponding Chern numbers in the zero-temperature limit, scaled by a factor of $1/D$ for systems with $D$-fold degenerate ground states. This proof establishes a precise correspondence between the topological invariants of pure and mixed states, addressing a longstanding challenge in the study of finite-temperature topological phases.

We have applied this framework to diverse systems, including a two-level system, a  coherent state model, a 2D Haldane model, and a four-band model, demonstrating the versatility and robustness of the thermal Uhlmann-Chern number. Numerical results for the Haldane and four-band models reveal the temperature-dependent behavior of these invariants, with topological signatures persisting at finite temperatures but diminishing as thermal fluctuations dominate. The thermal Uhlmann curvature, which reduces to the Berry curvature at zero temperature, provides further insight into the gradual loss of topological order with increasing temperature, as visualized across the Brillouin zone.

Our findings offer a new perspective on the interplay between topology and thermal effects, with potential implications for the experimental detection of topological phases in finite-temperature systems. Future research may explore the application of the thermal Uhlmann-Chern number to other topological systems, such as higher-dimensional or non-Hermitian models, and investigate its connections to observable quantities, such as transport properties. The mathematical framework established here lays a foundation for understanding topology in realistic quantum systems beyond the idealized zero-temperature limit, with potential impacts on fields ranging from fault-tolerant quantum computation to finite-temperature phase transitions in topological matter.


\section*{ACKNOWLEDGEMENTS}
H.G. was supported by the Innovation Program for Quantum Science and Technology (Grant No. 2021ZD0301904) and the National Natural Science Foundation of China (Grant No. 12074064). X.Y.H. was supported by the Jiangsu Funding Program for Excellent Postdoctoral Talent (Grant No. 2023ZB611).

\appendix

\section{$\text{Tr}\mathcal{F}_\text{U}=0$}\label{app1b}

By introducing the projection operator $P_i = | i \rangle \langle i |$ and defining $C_{ij} = \frac{(\sqrt{\lambda_i} - \sqrt{\lambda_j})^2}{\lambda_i + \lambda_j}$, the Uhlmann connection can be expressed as:
\begin{align}
	\mathcal{A}_{\mathrm{U}} &= - \sum_{i \neq j} C_{ij} P_i \mathrm{d} P_j.
\end{align}
The Uhlmann curvature is then given by:
\begin{align}
	\mathcal{F}_{\mathrm{U}} &= - \sum_{i \neq j} (\mathrm{d} C_{ij}) \wedge ( P_i \mathrm{d} P_j ) \notag \\
	&\quad + \sum_{i \neq j} \sum_{k \neq l} C_{ij} C_{kl} ( P_i \mathrm{d} P_k ) \wedge ( P_k \mathrm{d} P_l ). \label{eq:Uhlmann_F}
\end{align}
The trace of the first term, $(\mathrm{d} C_{ij}) \wedge ( P_i \mathrm{d} P_j )$, can be written as:
\begin{align}
	\sum_{i \neq j} (\mathrm{d} C_{ij}) \wedge \operatorname{Tr} ( P_i \mathrm{d} P_j ) &= \sum_{i \neq j} \sum_{m} (\mathrm{d} C_{ij}) \wedge \langle i | \mathrm{d} j \rangle \delta_{mi} \delta_{jm}. \label{eq:Uhlmann_F2}
\end{align}
This expression contributes only when $i = m = j$, but since the summation excludes $i = j$, the trace vanishes. Similarly, the trace of the second term in Eq.~\eqref{eq:Uhlmann_F} is zero, as evaluated by:
\begin{align}
	\operatorname{Tr} \left( \mathcal{A}_{\mathrm{U}} \wedge \mathcal{A}_{\mathrm{U}} \right) &= \sum_{i \neq j} C_{ij} C_{ji} \left\langle i | \partial_\mu j \right\rangle \left\langle j | \partial_\nu i \right\rangle \mathrm{d} R^\mu \wedge \mathrm{d} R^\nu \notag \\
	&= \sum_{i \neq j} C_{ji} C_{ij} \left\langle j | \partial_\nu i \right\rangle \left\langle i | \partial_\mu j \right\rangle \mathrm{d} R^\nu \wedge \mathrm{d} R^\mu \notag \\
	&= -\operatorname{Tr} \left( \mathcal{A}_{\mathrm{U}} \wedge \mathcal{A}_{\mathrm{U}} \right) = 0,
\end{align}
where the dummy indices $i \leftrightarrow j$ and $\mu \leftrightarrow \nu$ are exchanged in the second line. Consequently, $\operatorname{Tr} (\mathcal{F}_{\mathrm{U}}) = 0$. However, this result applies only to finite-dimensional systems. For infinite-dimensional systems, the trace of the Uhlmann curvature, $\operatorname{Tr} (\mathcal{F}_{\mathrm{U}})$, may not be well-defined. In subsequent sections, we illustrate this point by examining the infinite-dimensional harmonic oscillator coherent states as a specific example. For an even-dimensional system, the Chern number $\text{Ch}_{\mathrm{U}} = \frac{\mi}{2\pi} \int \operatorname{Tr} \mathcal{F}_{\mathrm{U}} = 0$, which aligns with the topological triviality of the Uhlmann bundle.

\bibliographystyle{apsrev}

\end{document}